\documentclass[prd,aps,twocolumn,showpacs,amsmath,amssymb]{revtex4}
\usepackage{amsmath} \usepackage{amssymb}
\usepackage{amsthm}
\usepackage{amsfonts}
\usepackage{hyphenat}
\usepackage{latexsym}
\usepackage{graphicx}
\usepackage{epsfig}
\newcommand{\field}[1]{\mathbb{#1}} 
\newtheorem{thm}{Theorem}
\newtheorem{cor}[thm]{Corollary}

\def\be{\begin{equation}} \def\ee{\end{equation}} \def\bea{\begin{eqnarray}} \def\eea{\end{eqnarray}}

\begin{document}


\title{Preheating with Non-Standard Kinetic Term}

\author{Jean Lachapelle and Robert H. Brandenberger} \affiliation{Department of Physics, McGill University, 3600
University Street, Montr\'eal QC, H3A 2T8, Canada}

\date{\today}


\begin{abstract}

We consider reheating after inflation in theories with a non-standard kinetic term. We show that the equation that exhibits parametric resonance is the Hill-Whittaker equation, which is a particular case of the more general Hill equation. This equation is in general more unstable than the Mathieu equation, such that narrow resonance preheating may be efficient in theories with a non-canonical kinetic term.
\end{abstract}

\maketitle

\section{Introduction} \label{sec:1}
Reheating is the period after inflation during which the energy stored in the inflaton is transferred to the matter and radiation of which the universe is made up. During this stage, the inflaton oscillates about its minimum and decays into relativistic particles. Through coupling with other fields, these oscillations can give rise to parametric instability and lead to an explosive growth in particle density. This phase, called \emph{preheating} \cite{TB,KLS1,STB,GKLS,KLS2}, plays a crucial role in the dynamics of reheating of an inflationary universe.

It has been shown \cite{BV,FB2} (see also \cite{FB,BKM}) that parametric
amplification of entropy modes during reheating can lead to a change in the
curvature power spectrum on scales larger than the Hubble radius
(but smaller than the horizon) without violating causality. Indeed entropy
perturbations act as a source for large scale curvature perturbations. The
effect is negligible for single-field inflation - where entropy perturbations
are suppressed on super-Hubble scales - but can be large in the case of
multiple field inflation \cite{GWBM}. The study of this mechanism in
multiple field models demands attention since it can significantly modify
the usual inflationary picture.

It has been shown \cite{Malik} that for the simplest preheating model,
the contribution to the large scale power spectrum of curvature perturbations
from entropy modes is negligible. However, very little work has been done to
this date to show that this is also the case for more complicated scenarios.
With the large number of inflationary models proposed today, one can expect
the study of this mechanism to become a powerful tool.

In this paper, we study preheating in two-field inflation models
$\{ \phi, \chi \}$ with a non-standard kinetic term for $\chi$. Such a
kinetic term arises inevitably in string-inspired models where the
K\"{a}hler potential $\mathcal{K}$ has a non-trivial geometry. Examples of
this are brane inflation models, where the inflaton is the distance between
two branes in higher dimensional spacetime, and modular inflation, where
inflation is caused by the dynamics of the deformations of a six dimensional
Calabi-Yau manifold (see e.g. \cite{Linde,Burgess,Cline,Tye,Eva} for reviews
on string inflation models).

Using the general form of a two-field Lagrangian with a non-standard kinetic term 
introduced in \cite{diMarco} (see also \cite{lalak}), we show that the entropy 
fluctuations during reheating obey the Hill-Whittaker equation, a modified version of 
the Mathieu equation (the equation that exhibits parametric instability). Indeed, the non-canonical part of the kinetic term enters as an oscillatory damping term into the equation, modifying its stability behaviour. We show that the Floquet theorem applies for this new type of equation such that the modes $\chi_k$ undergo parametric resonance for a wide range of parameters. Regions of instability appear as bands in the parameter space. We show that they are larger in the case  of the modified Mathieu equation. Therefore preheating is more likely to happen in models with non-standard kinetic terms.

\section{Preheating} \label{sec:2}

Consider a theory of two interacting scalar fields  with non-canonical kinetic term \cite{diMarco,lalak}
\begin{eqnarray}
\mathcal{S}=&\int d^4x \sqrt{-g} [ \frac{M^2_P}{2}R -\frac{1}{2} (\partial_{\mu} \phi)(\partial^{\mu} \phi) \nonumber  \\
 &- \frac{e^{2b(\phi)}}{2} (\partial_{\mu} \chi)(\partial^{\mu} \chi)-V(\phi,\chi)],
\end{eqnarray}
where $M_P$ is the reduced Planck mass and $b(\phi)$ is a non-trivial function of $\phi$ that renders the kinetic term non-canonical. This
type of action appears in the Roulette inflation model \cite{Roulette}.
It also appears in generalized Einstein theories
\cite{Starob1,Starob2}.
In a Friedmann-Robertson-Walker (FRW) background
\be
ds^2=-dt^2+a(t)^2d	 \textbf{x}^2,
\ee
with $a(t)$ denoting the scale factor, the equations of motion are
\begin{eqnarray}
\ddot{\phi}+3H\dot{\phi}+V_{\phi}=b_{\phi}e^{2b}\dot{\chi}^2 \\
\ddot{\chi}+(3H+2b_{\phi}\dot{\phi})\dot{\chi}+e^{-2b}V_{\chi}=0  \label{EOMCHI} \\
H^2=\frac{1}{3M_P^2} \left[ \frac{1}{2} \dot{\phi}^2+\frac{e^{2b}}{2}\dot{\chi}^2+V \right] \label{FRW1} \\
\dot{H}=-\frac{1}{2M_P^2} \left[\dot{\phi}^2+e^{2b}\dot{\chi}^2 \right]\label{FRW2}.
\end{eqnarray}
Note that for $b=0$ the equations of motion for the fields $\phi$ and $\chi$
reduce to the Klein-Gordon equations and Eqs.(\ref{FRW1})-(\ref{FRW2})
reduce to the well known Friedman equations in the presence of scalar
field matter.


Consider a model of chaotic inflation with a symmetry breaking potential
\be
V(\phi,\chi)=\frac{1}{2}m^2(\phi-\sigma)^2+\frac{1}{2}m_{\chi}^2\chi^2+\frac{1}{2}g^2\phi^2\chi^2
\ee
where $\phi$ is the inflaton and $\chi$ is the reheat field. We start with a
vanishing $\chi$ background field and will study the growth of the
perturbations $\delta \chi$ which begin as initial quantum vacuum
fluctuations. Inflation occurs for $|\phi-\sigma| \geq M_P$. In the
case of a standard kinetic term ($b=0$) and in the limit $m\gg m_{\chi}$,
this model gives efficient preheating \cite{TB,KLS2}. We shall
investigate the effect on this system of a non-vanishing $b(\phi)$.

After the shift $\phi-\sigma \rightarrow \phi$, the potential becomes
\begin{eqnarray}\label{potential}
V(\phi,\chi) \, &=& \, \frac{1}{2}m^2\phi^2 + \frac{1}{2}m_{\chi}^2\chi^2 +
\frac{1}{2}g^2\phi^2\chi^2 + g^2\phi\sigma\chi^2  \nonumber \\
&& +\frac{1}{2}g^2\sigma^2\chi^2 \, .
\end{eqnarray}
When $\phi \simeq M_P$, inflation ends and the inflaton starts oscillating
about the minimum of its potential
\be
\phi(t) \, = \, \Phi(t)cos(mt) \, ,
\ee
where $\Phi(t)$ is the amplitude of oscillation. This amplitude decreases
as $\sim 1/t$ due to Hubble friction. In the case when the period of
oscillations is small compared to the Hubble time, i.e. $m \gg H$, we can neglect
the Hubble friction as a first approximation: preheating happens on a
time scale much shorter than the Hubble scale and we expect other effects -
such as back-reaction from the newly created $\chi$ particles - to come
into play long before Hubble friction is felt.

Consider small amplitude oscillations $\Phi \ll \sigma$, such that the term
$\frac{1}{2}g^2\phi^2\chi^2$ in the potential can be neglected. Then,
the equation for perturbations $\delta \chi$ is
\be\label{perturbedEOM}
\ddot{\delta\chi}+2\dot{b}\dot{\delta \chi}+\frac{k^2}{a^2}+e^{-2b}\left[g^2\sigma^2+m_{\chi}^2+2g^2\sigma\Phi cos(mt) \right]\delta \chi=0
\ee
To first order in $\left( \frac{\Phi}{\sigma}\right)$, the oscillatory
friction term is
\be
\dot{b} \, = \, -b_{(\phi/\sigma)}^o \left( \frac{\Phi}{\sigma}\right) m sin(mt)
\ee
and the prefactor of the last term on the left hand side of (\ref{perturbedEOM})
takes the form
\be
e^{-2b} \, = \, e^{-2b_o} \left(1-2b_{(\phi/\sigma)}^o
\left( \frac{\Phi}{\sigma} \right) cos(mt) \right),
\ee
where
$b^o_{(\phi/\sigma)} \equiv \frac{db}{d(\phi/\sigma)} \Big |_{\phi=0}$
and
$b_o \equiv b \Big |_{\phi=0}$.

Performing the change of variables $mt \rightarrow 2z$, and neglecting
$\mathcal{O}(\left(\frac{\Phi}{\sigma}\right)^2)$ contributions, we get
\begin{equation}\label{modmat}
\delta\chi'' + 2p \cdot sin(2z)\delta\chi'  + [A_k-2qcos(2z)]\delta\chi
\, = \, 0 \, ,
\end{equation}
where a prime denotes the derivative with respect to $z$ and
\begin{eqnarray}
A_k \, &=& \,
\frac{4k^2}{m^2a^2}+\frac{4e^{-2b_o}}{m^2}\left(m_{\chi}^2+g^2\sigma^2\right) \\
q \, &=& \, \frac{4e^{-2b_o}}{m^2}
\left(g^2 \sigma^2\left( b^o_{(\phi/\sigma)}-1 \right)+m_{\chi}^2b^o_{(\phi/\sigma)} \right)
\left( \frac{\Phi}{\sigma}\right) \nonumber \\
p \, &=& \, -2 b^o_{(\phi/\sigma)} \left( \frac{\Phi}{\sigma} \right) \, .
\nonumber
\end{eqnarray}
Eq. (\ref{perturbedEOM}) is reminiscent of the well known Mathieu equation,
but with an additional oscillatory friction/anti-friction
term with coefficient $2p$. Defining the new function 
$f(z)=e^{-p\cos(2z)/2} \delta \chi$, 
Eq.(\ref{modmat}) gives
\begin{equation}
f''+\left[A_k-
\frac{p^2}{2}-2(q+p)\cos (2z)+\frac{p^2}{2} \cos(4z)\right]f=0,
\end{equation}
which is known as the Whittaker-Hill equation and is a particular case of Hill's equation \cite{magnuswinkler,urwinarscott}. The Floquet theorem applies for this type of equation 
and implies that any solution can be written in the form
\begin{equation}\label{whittaker}
f(z)=\alpha e^{\mu z}\phi(z)+\beta e^{-\mu z}\phi(-z),
\end{equation}
where $\mu \in \field{C}$ is called the Floquet exponent, $\phi(z)$ is periodic in $z$ with 
period $\pi$, and $\alpha$, $\beta$ are real constants. If $\mu$ is imaginary, the solution 
is stable whereas $\mu$ real gives exponential instability. Note that the Floquet exponent 
$\mu$ - which determines the rate of exponential growth of solutions - is the same for 
$\delta \chi(z)$ and $f(z)$.

\section{Stability of the Whittaker-Hill equation}

The Floquet theorem implies that the Whittaker-Hill equation can exhibit exponential 
instability. A method for calculating the resonance was derived by Whittaker in \cite{whittaker} 
for the full Hill equation, which includes the Whittaker-Hill equation as a particular case. 
For clarity, we present the details of this method. We start by writing a solution of the form
\begin{eqnarray}
y(z) \, &=& \, e^{\mu z}\phi(z) \,
= \, e^{\mu z} \sum_{-\infty}^{\infty} c_{2r}e^{2riz} \nonumber \\
&=& \, \sum_{-\infty}^{\infty} c_{2r}e^{(\mu+2ri)z} \, ,
\end{eqnarray}
and plug it into Eq.(\ref{whittaker}).
We obtain the following recursion relation for the coefficients:
\begin{align}
&\left(c_{2(r+2)}+c_{2(r-2)}\right)\frac{p^2/4}{(i\mu-2r)^2-a} \nonumber \\  &
+\left(c_{2(r-1)}+c_{2(r+1)}\right)\frac{q+p}{(i\mu-2r)^2-a}+c_{2r}=0
\end{align}
where we use the abbreviation $a$ for $A_k$, which can be written as
\begin{equation}
c_{2(r-2)}\xi_{2r}+c_{2(r-1)}\gamma_{2r}+c_{2r}+c_{2(r+1)}\gamma_{2r}+c_{2(r+2)}\xi_{2r}=0,
\end{equation}
where
\begin{align}
\gamma_{2r} & \equiv \frac{q+p}{(i\mu-2r)^2-a}, & \xi_{2r} \equiv &\frac{p^2/4}{(i\mu-2r)^2-a}.
\end{align}
The value of $i\mu$ which solves this recursion relation is that for which the determinant of the coefficients vanishes
\[ \Delta(i\mu) \equiv \left| \begin{array}{ c c c c c c c}
...  & & &  & \\
\xi_{-2} & \gamma_{-2} & 1 &\gamma_{-2}& \xi_{-2} & & \\
&\xi_{0} & \gamma_{0} & 1 &\gamma_{0} &\xi_{0} & \\
 &  & \xi_{2} &\gamma_{2} & 1 &\gamma_{2}&\xi_{2}  \\
 & & & & & ...
\end{array} \right| =0. \]
Let us investigate the singularities of this determinant in the complex plane. For $\sqrt{a}$ not an integer, the determinant has simple poles at
\begin{equation}
i\mu=2r\pm\sqrt{a}.
\end{equation}
There are pairs of poles at distance $\sqrt{a}$ of every point $2l$ on the real line ($l$ a real integer). 
Let us show that the residues at these poles alternate in sign and are periodic. This will allow us 
to construct an entire function $\zeta(z)$ which will help us solve the recursion relation.
\begin{thm}\label{importthm}
\begin{align}
(i) &&&  \text{$\Delta(i\mu)$ is periodic in the reals with period $2m$,} \notag \\
&&& \text{where $m$ is a real integer,} \notag \\
(ii) &&& \text{$\Delta(i\mu)$ is even in the reals about every point $2m$}.
\end{align}
\end{thm}
\begin{proof}
One can check that the shift $i\mu \rightarrow i\mu +2m$, ($m$ a real integer) induces the change
\begin{align}
\gamma_{2r} &\rightarrow \gamma_{2(r-m)} \nonumber \\
\xi_{2r} &\rightarrow \xi_{2(r-m)}
\end{align}
such that the rows of the determinant are all shifted up by an equal amount. This does not change 
the value of the determinant of the infinite matrix. To prove $(ii)$, one can check that the 
change $i\mu \rightarrow -i\mu$ induces the shift
\begin{align}
\gamma_{2r} \rightarrow \gamma_{-2r} \nonumber \\
\xi_{2r} \rightarrow \xi_{-2r}
\end{align}
such that the rows of the determinant are flipped about the row with the $\gamma_0$'s, i.e.

\[ \Delta(i\mu) \equiv \left| \begin{array}{ c c c c c c c}
...  & & &  & \\
\xi_{2} & \gamma_{2} & 1 &\gamma_{2}& \xi_{2} & & \\
&\xi_{0} & \gamma_{0} & 1 &\gamma_{0} &\xi_{0} & \\
 &  & \xi_{-2} &\gamma_{-2} & 1 &\gamma_{-2}&\xi_{-2}  \\
 & & & & & ...
\end{array} \right| =0. \]
From the method of cofactors one can check that this is equal to $\Delta(i\mu)$.
\end{proof}
\begin{cor}
For any even integers $r_1$ $r_2$, the residues of $\Delta(i\mu)$ satisfy
\begin{align}
(i) &&& Res(\Delta(i\mu),r_{1} + \sqrt{a})=Res(\Delta(i\mu),r_{2} + \sqrt{a}), \notag \\
(ii) &&& Res(\Delta(i\mu),r_{1} - \sqrt{a})=Res(\Delta(i\mu),r_{2} - \sqrt{a}), \notag \\
(iii) &&& Res(\Delta(i\mu),r_1 + \sqrt{a})=-Res(\Delta(i\mu),r_1 - \sqrt{a}),
\end{align}
where $Res(\Delta(i\mu),z)$ stands for the residue of $\Delta(i\mu)$ at point $z$.
\end{cor}
\begin{proof}
The proof follows directly from Theorem \ref{importthm}. Let us perform a Laurent expansion about each of the poles. In some neighbourhood around $i\mu=\sqrt{a}$, we have
\begin{equation}
\Delta(i\mu)=\frac{C_{-1}}{(i\mu-\sqrt{a})}+C_0+C_1(i\mu-\sqrt{a})+...
\end{equation}
whereas, in some neighborhood around $i\mu=-\sqrt{a}$, we have
\begin{equation}
\Delta(i\mu)=\frac{D_{-1}}{(i\mu+\sqrt{a})}+D_0+D_1(i\mu+\sqrt{a})+...
\end{equation}
We see that $\Delta(i\mu)$ is even if and only if $C_{-1}=-D_{-1}$, $C_{0}=D_{0}$, $C_{1}=-D_{1}$... The constants $C_{-1}$ and $D_{-1}$ are the residues at the poles.
\end{proof}

Let us introduce a function $\lambda$ which has the same analyticity structure as our determinant.
\begin{equation}
\lambda(i\mu) = \frac{1}{cos(\pi i \mu)-cos(\pi \sqrt{a})}.
\end{equation}
This function is periodic $i\mu \rightarrow i\mu+2m $ and has simple poles at $i\mu=2r \pm \sqrt{a}$, just like $\Delta(i\mu)$. From the periodicity of $\lambda$, we conclude that the residues
$Res(\lambda,2r + \sqrt{a})$ are the same for all $r$ and similarly for $Res(\lambda,2r - \sqrt{a})$. Left to prove that the residues alternate in sign, i.e.
\begin{equation}
Res(\lambda,2r + \sqrt{a})=-Res(\lambda,2r - \sqrt{a}).
\end{equation}
From l'Hopital's rule, the residues can be calculated explicitly
\begin{equation}
\lim_{i\mu \to 2r \pm \sqrt{a}} (i\mu -r \mp \sqrt{a})\lambda  = \mp\frac{1}{\pi sin(\pi \sqrt{a})},
\end{equation}
which shows that the sign of the residue is switched between the two types of poles. The poles of the function $\lambda$ match the poles of $\Delta(i\mu)$ up to a constant. Thus an entire (analytic everywhere) function can be constructed
\begin{equation}\label{zeta}
\zeta=\Delta(i\mu)-\kappa\lambda,
\end{equation}
where $\kappa$ is a constant set to cancel exactly the divergences at the singular points. The functions $\lambda$ and $\Delta(i\mu)$ have limiting behavior
\begin{align}
\lim_{Im(i\mu) \to\infty}\lambda =& 0, \\
\lim_{|i\mu| \to\infty}\Delta(i\mu)=&1,
\end{align}
since $\gamma_{r}$ goes to $0$ as $|i\mu|\longrightarrow \infty$ (in which case $\Delta(i\mu)$ approaches the identity matrix). Therefore,
\begin{equation}
\lim_{Im(i\mu) \to\infty}\zeta(i\mu)=1.
\end{equation}
This implies that $\zeta$ is bounded. From Liouville's theorem an entire function whose absolute value is bounded throughout the $z$-plane is constant. Thus $\zeta=1$ throughout the z-plane, in particular $\zeta(0)=1$. From this we get
\be
\zeta(0)=1=\Delta(0)-\kappa \lambda(0),
\ee
such that
\be
\kappa=\frac{\Delta(0)-1}{\lambda(0)}=(\Delta(0)-1)(1-cos(\pi\sqrt{a})).
\ee
From (\ref{zeta}), we have
\begin{equation}
\Delta(i\mu)=1+\frac{(\Delta(0)-1)(1-cos(\pi\sqrt{a}))}{cos(\pi i\mu)-cos(\pi \sqrt{a})}
\end{equation}
The determinant equation is $\Delta(i\mu)=0$. It is solved for
\begin{equation}
cos(\pi i \mu)=1+\Delta(0)[cos(\pi\sqrt{a})-1].
\end{equation}
The expression for $\mu$ (which is derived by Whittaker in \cite{whittaker}) follows
\begin{equation}\label{muformula}
\mu=\frac{-i}{\pi}cos^{-1} \left\{ 1+\Delta(0)[cos(\pi\sqrt{a})-1] \right\}.
\end{equation}
This expression has the advantage that it can be evaluated efficiently
using numerical methods. Indeed, the quantity $\Delta(0)$ is easily
approximated by taking numerically the determinant of an $n\times n$
matrix with large $n$ (the center of the matrix is at the position of the 
$1$ in the row containing the $\gamma_0$). 
Since the off-diagonal elements all scale as $1/r$
as we go away from the center of the matrix, there will be a rapid
convergence of the determinant computed in this approximation.

\begin{figure} [h]
\centering
\begin{tabular}{c}
\epsfig{file=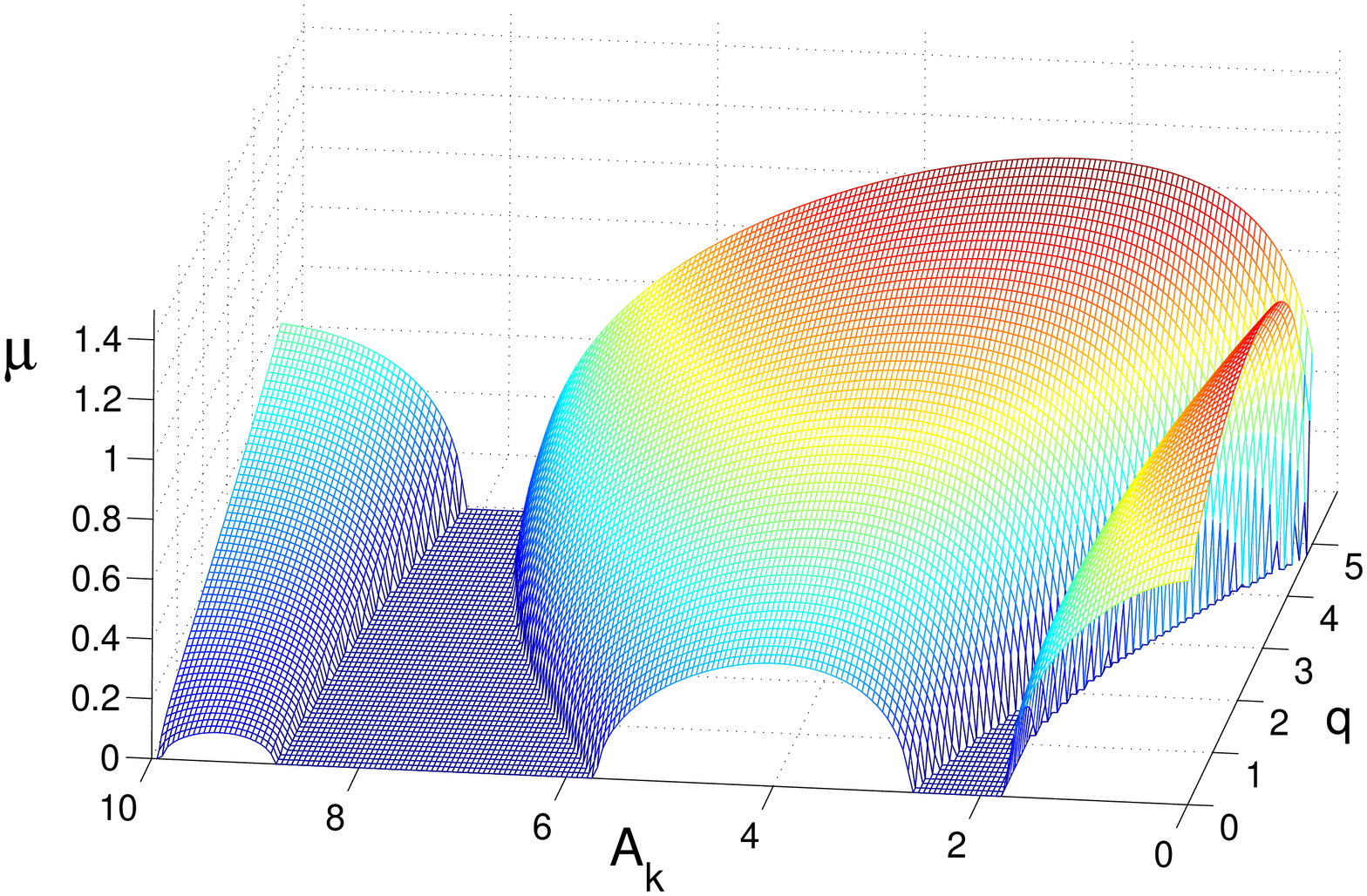,width=0.8\linewidth,clip=} \\
\epsfig{file=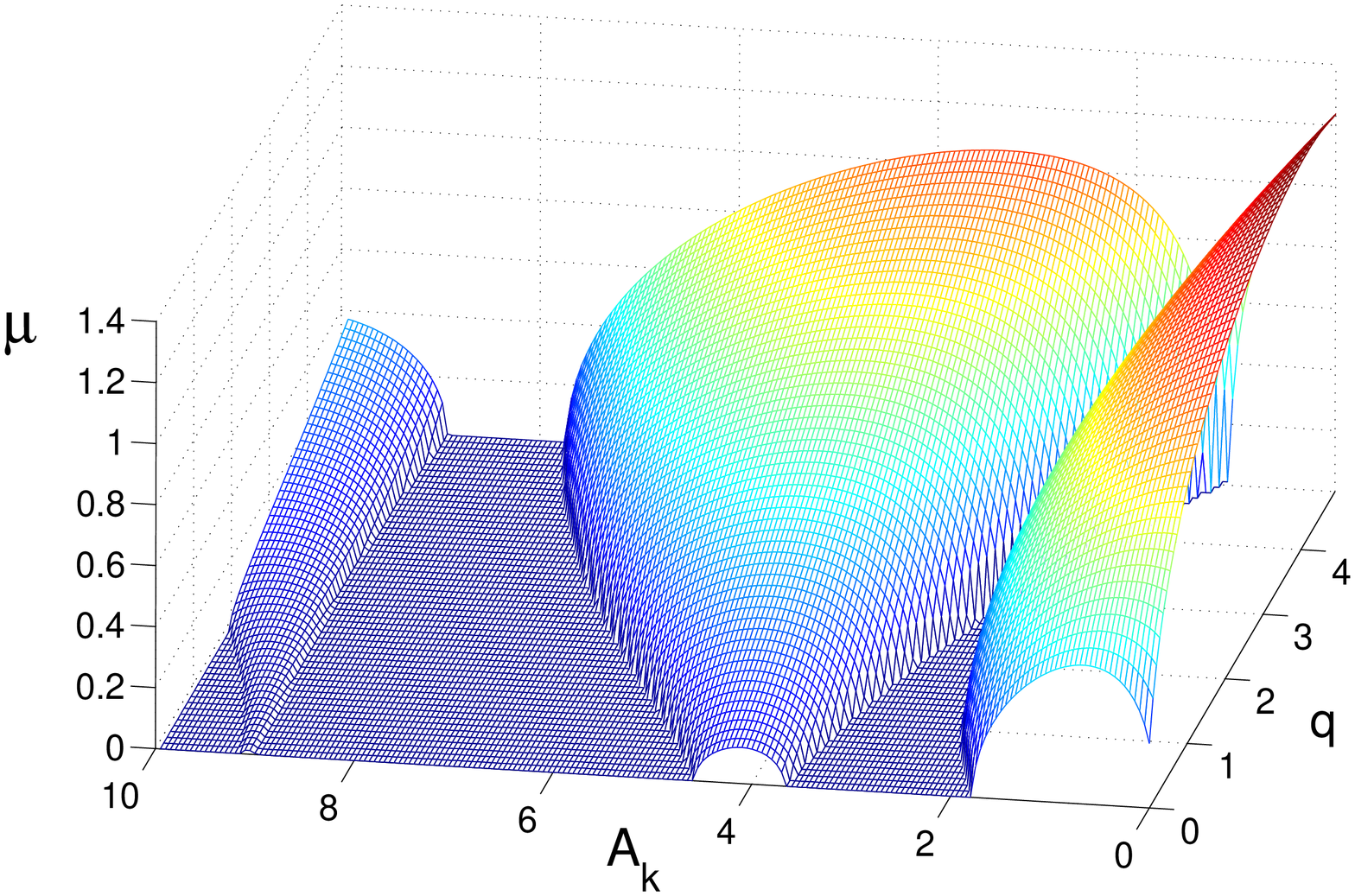,width=0.8\linewidth,clip=} \\
\epsfig{file=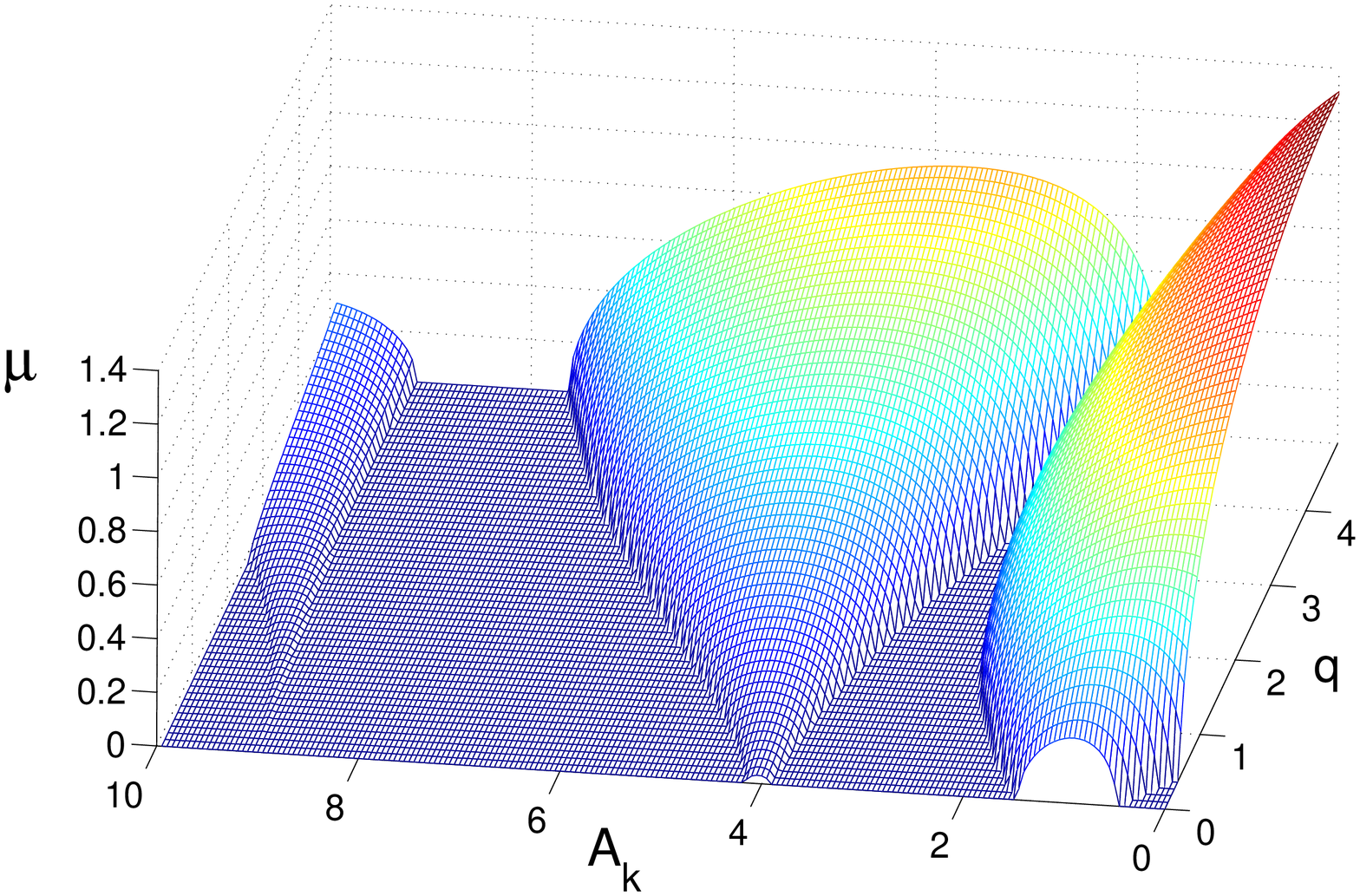,width=0.8\linewidth,clip=} \\
\epsfig{file=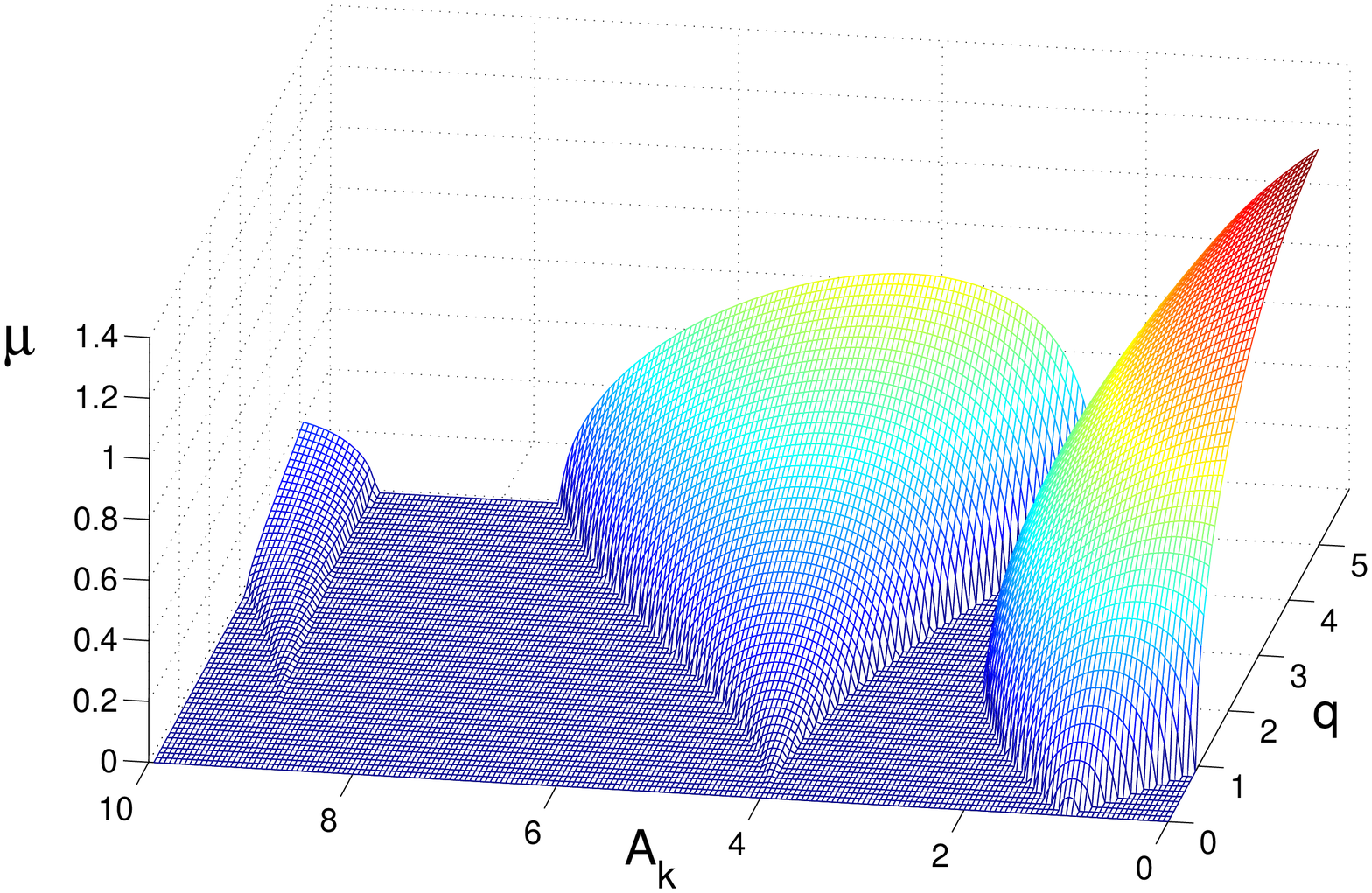,width=0.8\linewidth,clip=} \\
\end{tabular}
\caption{ The Floquet exponent $\mu$ for (from top to bottom) $p=2$, $p=1$, $p=0.5$ and $p=0$} \label{pis}
\end{figure}

Figure \ref{pis} show the real part of the Floquet exponent
$\mu$ in the $(a,q)$ parameter space for different values of $p$. The numerical simulations were performed using a $500 \times 500$ matrix The instability bands of the $p=0$ figure match 
exactly those of the well-known Mathieu equation, which demonstrates the accuracy of the procedure.
The first observation is that
for bigger values of $p$, the instability bands get larger. This
indicates that inflationary models with non-standard kinetic terms can
exhibit more efficient parametric resonance.
The heuristic reason for the more efficient resonance is the fact
that the periodic variation in the friction/anti-friction term in
the equation of motion contains energy which can be given to the
system. For a more mathematical argument, let us return to the
expression (\ref{muformula}) for the Floquet exponent $\mu$. If
we increase the value of $p$, then the value of the determinant
$\Delta(0)$ will increase. This will then lead to an increase in
the absolute value of $\mu$.

Next, we note the rich structure of the parameter space. The symmetry about
the $q=0$ axis that appears for the usual Mathieu equation is broken when
$p$ acquires a non-vanishing value. Moreover, the resonance bands are
shifted from their usual positions ($A_k=1,4,9,...$) as they become larger.

\section{Potential Without Symmetry Breaking}\label{nosymbreak}

Consider the simple potential for chaotic inflation with a massless $\chi$ field
\be
V=\frac{1}{2}m^2\phi^2+\frac{1}{2}g^2\phi^2\chi^2.
\ee
For this type of potential, the condition for efficient
preheating (with $m \gg H$  and in the absence of non-standard kinetic term)
is \cite{KLS2}
\be\label{condition}
q^2m = \frac{g^4\Phi^4}{16m^3} \ge H.
\ee
If condition (\ref{condition}) is not satisfied, the modes are red-shifted outside the resonance band before they have time to get amplified and preheating does not takes place. However, this condition translates to
\be
\Phi \, \ge \, (\frac{m}{M_P})^{1/3} (\frac{2}{g})^{4/3} m \, .
\ee
Since $m$ must be much smaller than the Planck mass, but $\phi$
starts oscillating with a value close to the Planck mass, this
condition is easy to satisfy.

We show that by introducing non-vanishing $b$ we can relax this condition. Consider small oscillations $\phi \ll M_P$ about the minimum. We expand
\be
e^{-2b}=e^{-2b_o}\left(1-2b_{(\phi/M_P)}^o \left(\frac{\phi}{M_P} \right) \right),
\ee
where $b_{(\phi/M_P)}\equiv \frac{db}{d(\Phi/M_P)} \Big |_{\phi=0}$ such that the equation for the perturbations becomes
\begin{align}\label{expand}
&\delta \chi''-4b^o_{(\phi/M_P)} \left( \frac{\Phi}{M_P} \right) sin(2z)\delta \chi' \nonumber \\
&+\{ 2e^{-2b_o}\frac{g^2\Phi^2}{m^2}\left[1-3b^o_{(\phi/M_P)}\left( \frac{\Phi}{M_P} \right) cos(2z)+cos(4z)\right] \nonumber \\
& +\frac{4k^2}{a^2m^2} \}\delta \chi=0,
\end{align}
after dropping the higher frequency contribution proportional to $cos(6z)$. Two cases are of 
interest: first, if $b^o_{(\phi/M_P)}\left(\frac{\Phi}{M_P} \right) \ll 1$, the damping term becomes 
negligible and $cos(4z)$ dominates, such that the equation can be turned into a standard 
Mathieu equation after  the change of variables $2z \rightarrow z$. In the condition $q^2m \ge H$ 
for efficient resonance, we now have $q= e^{-2b_o}\frac{g^2\Phi^2}{2m^2}$. This case is 
realized in the Roulette Inflation model, which is discussed in a companion paper. Since
$m$ is much smaller than $M_p$, the efficient resonance condition is satisfied.

The second case is $b^o_{(\phi/M_P)}\left(\frac{\Phi}{M_P}\right) = \mathcal{O}(1)$ or greater. In this 
case the higher frequency contribution $ cos(4z)$ in the mass term can be neglected as a first approximation. We consider the situation $\frac{g^4\Phi^4}{m^3} <H $, such that the resonance is 
inefficient for $b=0$.

\begin{figure}[h]
\centering
\includegraphics[width=0.5\textwidth]{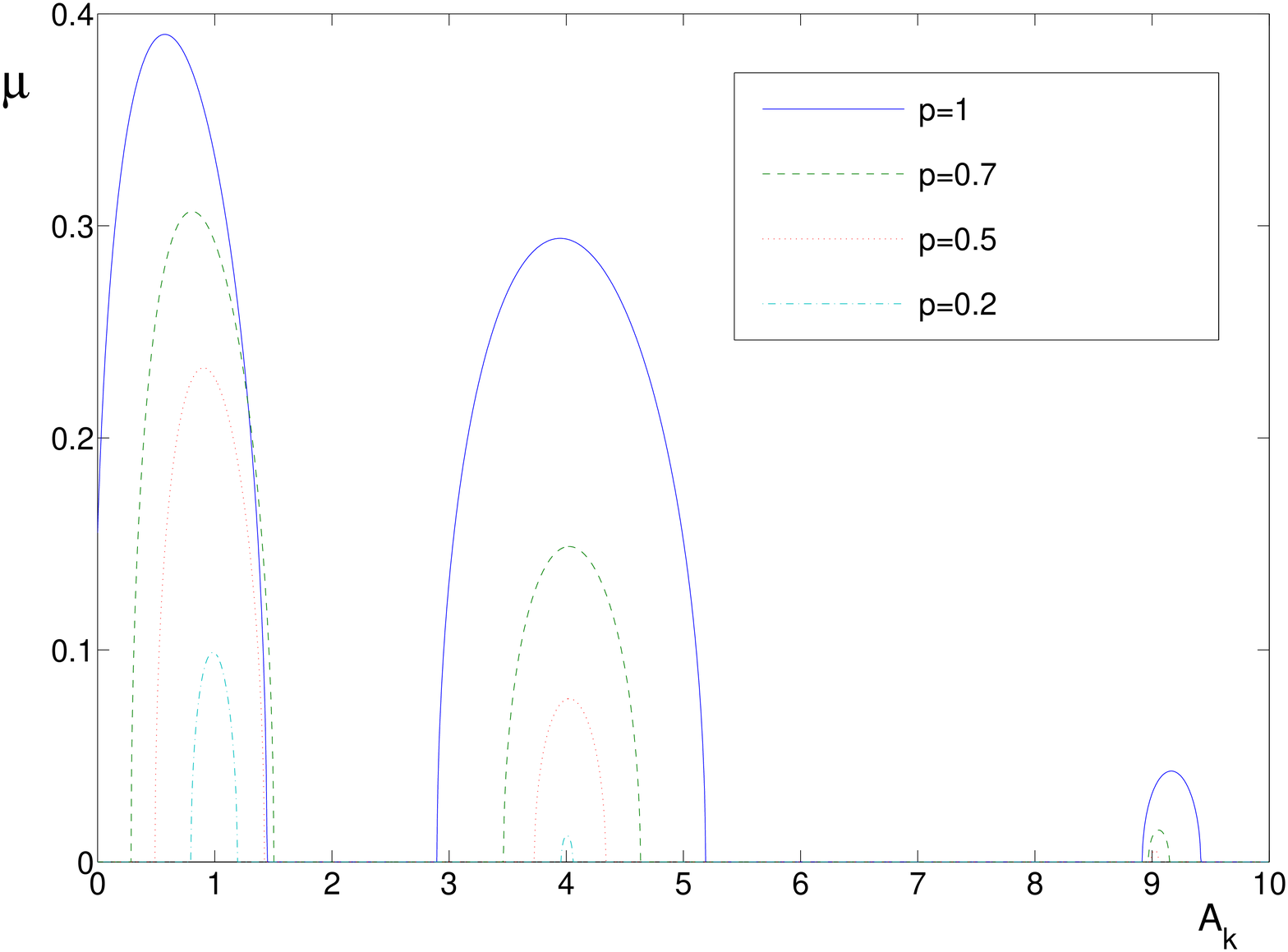}
\caption{The Floquet exponent $\mu$ on the $q=0$ axis for different values of $p$. }\label{reheatgraph}
\end{figure}

Consider $e^{-b_o}$ of order unity or smaller. We show that preheating can still happen in this 
case. First, note that $p$ dominates over $q$:
\be
\delta \chi''-4b^o_{(\phi/M_P)}\left(\frac{\Phi}{M_P}\right)sin(2z)\delta \chi'+\frac{4k^2}{m^2a^2}\delta \chi=0.
\ee
such that as a first approximation, the parameters of the modified Mathieu
equation can be written in the form
\begin{align}
&A_k=\frac{4k^2}{m^2a^2} \\
&q=0 \\
&p=-2b^o_{(\phi/M_P)}\left(\frac{\Phi}{M_P}\right).
\end{align}
The graphs of Fig. \ref{pis} shows that the modified Mathieu equation exhibits resonant behavior on the $q=0$ axis for $p$ non-vanishing, indicating that efficient resonance may be possible. Figure \ref{reheatgraph} shows the Floquet exponent on the $q=0$ axis for different value of $p$.

\begin{figure}[h]
\centering
\includegraphics[width=0.5\textwidth]{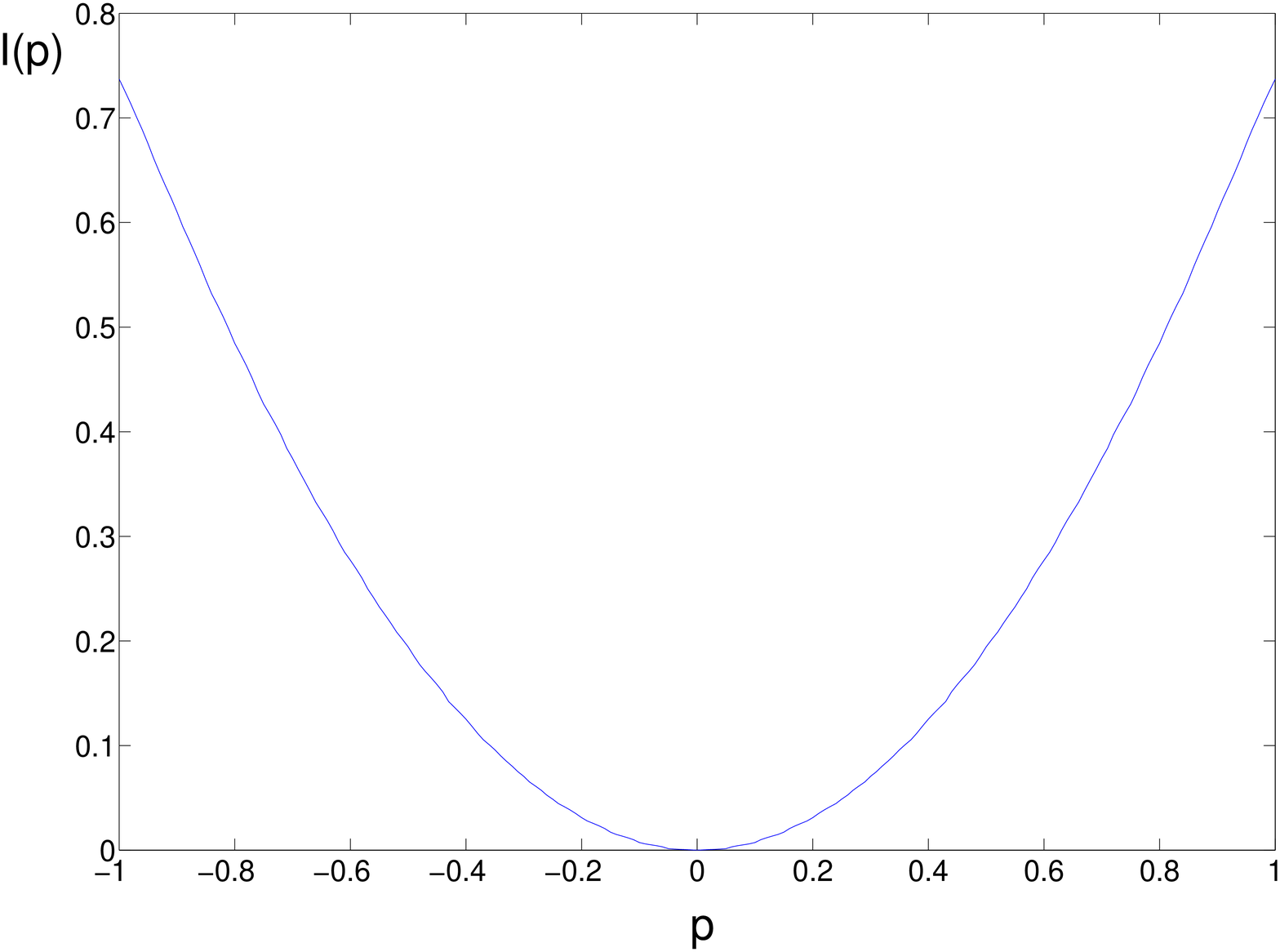}
\caption{Growth factor $I$ as a function of $p$. }\label{parabola}
\end{figure}

One mechanism which can prevent resonance from being efficient is the redshift of the modes outside the resonance band before they get sufficiently amplified. While in the band, the growth is
\be
\delta \chi \simeq \delta \chi_o e^{\frac{1}{2}\mu m \Delta t},
\ee
where $\delta \chi_o$ is the initial value of $\delta \chi$ and $\Delta t$ is a small time interval. Summing over the time intervals and taking the continuity limit gives
\be\label{integral}
\delta \chi \simeq \delta \chi_o e^{\frac{m}{2}\int \mu  dt},
\ee
with $\mu=\mu(t)$. The condition $m \gg H$ implies that $H\simeq$ constant on time scales relevant for preheating. The growth (\ref{integral}) can be integrated on the $A_k$ by performing the change of variables $dt=dA_k/H$. It becomes
\be\label{integral2}
\delta \chi \simeq \delta \chi_o e^{\frac{m}{2H}\int \mu  dA_k}.
\ee
For the growth to be effective for some mode $k$, we need
\be\label{int2}
\frac{m}{2H}\int \mu(A_k)  dA_k  \equiv \frac{m}{2H}I >1.
\ee
The integral $I$, which is the area under the curve shown in
Fig. \ref{reheatgraph}, is an even function of $p$ (we computed
the integral in the approximation of setting the factor $A_k^{-1}$
in the integrand to $1$ which is a good approximation since it
corresponds to the center of the resonance band). Plotting $I$
for $p\in [-1,1]$ shows that $I(p)$ can be approximated by a parabola
on this interval (see Fig. \ref{parabola}).

A least square fit gives
\be
I(p) \simeq 0.75 p^2.
\ee
During reheating $H \simeq \frac{\Phi m}{M_P\sqrt{6}}$ and condition ($\ref{int2}$) becomes
\be
\left(\frac{\Phi}{M_P} \right) \left( b_{(\phi/M_P)}^o\right)^2  >  \frac{1}{1.5\sqrt{6}}.
\ee
This condition does not contradict our first assumption $\left(\frac{\Phi}{M_P}\right)b_{(\phi/M_P)}^o \simeq \mathcal{O}(1)$ for small amplitude oscillations. This implies that preheating is efficient in this case.

Efficient parametric resonance is obtained in this chaotic inflation model for a broad range of parameters. Indeed, the introduction of a non-standard kinetic term that satisfies
 \begin{align}
 e^{-b_o} &\leq 1 \\
  b^o_{(\phi/M_p)} &\geq \left(\frac{M_P}{\Phi}\right)
 \end{align}
 gives rise to preheating whatever the value of $g^2\Phi^2/m^2$. This simple example teaches us that reheating in theories with non-standard kinetic term must be approached with caution since parametric resonance may happen in exotic ways.

 Those results hold for small amplitude oscillations. No direct comparison with the standard large field reheating results are possible at this point, since the expansion (\ref{expand}) becomes inaccurate for large amplitude oscillations. Note that in the case we studied here, the condition $q>1$ is easily satisfied in general and therefore we should expect to see broad resonance. The precise effect of a non-standard kinetic terms during the phase of broad resonance have yet to be investigated.

\section{Conclusions and discussions}

In this paper, we have studied the simplest preheating model with
a non-standard kinetic term. We observed that the non-standard kinetic 
term introduces an 
oscillating friction term in the Mathieu equation that modifies the
stability behavior of the perturbations in the reheat field $\chi$. 
The equation that is obtained 
is the Hill-Whittaker equation, which in the form in which it appears 
here is more unstable than the Mathieu equation.

Thus parametric resonance is more likely to be efficient in theories with a
non-standard kinetic term. In Section \ref{nosymbreak}, we showed that
for the simplest preheating model, there exists a function $b(\phi)$ that
makes parametric resonance efficient whatever the coupling $g$. Thus one
should be careful when dealing with theories with non-standard kinetic
terms in the context of reheating as models that do not seem to give rise
to preheating at first sight (because they have a small value of $q$)
can become resonant due to
the oscillatory friction/anti-friction term introduced by the non-standard
kinetic term in the action. Note, however, that for small values of $b(\phi)$,
as is realized in the Roulette Inflation model, the effects of the
non-standard kinetic term are negligible.

As discussed in a companion paper, equations like those studied
in this paper occur in the Roulette Inflation
model. There, the perturbations in the reheat field $\chi$ correspond to
entropy fluctuations that are parametrically amplified through the mechanism
discussed here. These entropy modes seed curvature perturbations on large
scales, altering homogeneity and scale invariance. In a companion paper, we
were able to put constraints on the parameters of the Roulette Inflation
model so as to match the results from the COBE experiment. Similar work was
conducted in the cases of the $D3/D7$ brane inflation model
\cite{BDD} and the KKLMMT model \cite{BFL}. In both
cases it was found that under certain assumptions, during reheating
entropy fluctuations can seed a secondary curvature mode that dominates
over the primary mode (the purely adiabatic linear perturbation theory mode).

Our results show that the study of parametric resonance during reheating can
be used to put constraints on the many inflationary models proposed today.
We expect the power of this method to be fully appreciated in the future.

\begin{acknowledgements}

This work is supported in part by an NSERC Discovery Grant, by the Canada Research Chair program, and by a FQRNT
Team Grant. We wish to thank Francis Cyr-Racine for valuable conversations.

\end{acknowledgements}

\bibliographystyle{unsrt}
\bibliography{bibl3}

\end{document}